\documentclass{article}
\usepackage{amsfonts}
\usepackage{amssymb}
\usepackage{amsmath}
\usepackage{graphicx}
\usepackage{color}
\usepackage{hyperref}

\usepackage{pstricks,pst-node,pst-text,pst-3d}
\newrgbcolor{Aqua}{1 0 1}

\begin{document}
\title{Parameter Estimation with Reluctant \\
Quantum Walks: \\
a Maximum Likelihood approach.}
\author{Demosthenes Ellinas$^{\diamond }$ Peter D. Jarvis$^{\ast }$ and
Matthew Pearce$^{\ast \# }$ \\
$^{\diamond }$Technical University of Crete \ School of ECE QLab \\
Chania Crete Greece \\
$^{\ast }$School of Natural Sciences, University of Tasmania\\
Hobart Tasmania Australia  \\
$^{\# }$current address: School of Physics and Astronomy, Monash University\\
Clayton Victoria Australia   \\
$^{\diamond }$\texttt{dellinas@tuc.gr} $^{\ast }$\texttt{%
Peter.Jarvis@utas.edu.au} \\$^{\ast }$\texttt{mpearce5@utas.edu.au}}
\maketitle

\begin{abstract}
The parametric maximum likelihood estimation problem is
addressed in the context of quantum walk theory 
for quantum walks on the lattice of integers.
A coin action is presented, with the real parameter $\theta$ to be estimated identified with the angular argument of an orthogonal reshuffling matrix. 
We provide analytic results for the probability distribution for a quantum walker to be displaced by $d$ units from its initial position after $k$ steps. For $k$ large, we show that the likelihood is sharply peaked at a  displacement determined by the ratio $d/k$\, which is correlated with the reshuffling parameter $\theta$. We suggest that this `reluctant walker' behaviour provides the framework for maximum likelihood estimation analysis, allowing for robust parameter estimation of $\theta$ via return probabilities of closed evolution loops and quantum measurements of the position of quantum walker  with`reluctance index' $r=d/k$. 
\end{abstract}
\pagebreak


\section{Introduction}

Quantum estimation theory \cite{helstrom},\cite{holevo},\cite{bc},\cite{bcm}, 
is developing as an important component of quantum information and
computation theory, which extends the concepts and techniques of the classical
theory into the quantum framework; see early \cite{reviewMLE}, \cite{MLE} and recent reviews: \cite{review1},\cite{review2},\cite{review3};   
The classical Fisher information techniques and the classical Cramer-Rao inequality have found various versions of quantum counterparts (see e.g. the last three reviews above). The quantum estimation theory plays a central role in the general framework of quantum technology, see e.g. \cite{milburn}.
The maximum likelihood (ML) principle concerns especially this work as a technique that would for example provide
parameter estimation in the form of operational valued functions,
the measurement of which would yield estimates for interesting parameters of
various quantum probability mass  functions (pmf). 
Within this general framework we address the problem
of the estimation of an angular variable that determines the coin matrix of a
one dimensional quantum walk QW (\cite{kempe} --\cite{kendon}), on the lattice of integers by the ML estimation method (MLE). 

Previous works have also addressed similar parameter estimation problems within the context of QWs see e.g. \cite{Laflamme2008coin,Paris2019probe,Paris2022metrology}),
wherein the quantum Fisher information and the quantum Cramer-Rao inequality have been used as tools in the various estimation task. In some cases also problems of multi-parameter estimation problems have been addressed.  
In those works the generic behaviour of the walker itself and the ensuing estimation questions are typically captured using 
numerical simulations. We note however that there is a need for obtaining analytic and exact solutions for quantum estimation problems in general models and in QWs in  particular. In addition to exact solutions it is also desirable to have estimation problems formulated in the language of quantum information and computation theory i.e. in such ways that concepts such as quantum channels, unitary dilations, quantum measurements as well as measures of resources e.g. entanglement or other general resources are used and utilized extensively.
With such broad aims in mind we address in this work the estimation problem 
mention previously by using a QW framework and the technique of maximum likelihood estimation (MLE) in the quantum context. The QW-MLE problem formulated is solved by
a novel exact closed form analytic solution for any finite number of steps for the 
pmf of the problem of QW on the integers with an arbitrary
orthogonal coin matrix. The analytic likelihood function constructed from
that pmf is shown to be sharply peaked  and so it allows for an in-principle exact parameter estimation. These results are further elaborated to provide an operational implementation of the estimation procedure in terms of quantum
observables, quantum measurements and CP maps and their
unitarily dilated equivalent forms. Finally a analysis of the computational aspects of computing the quantum likelihood function from data identified by quantum measurements of positional probabilities of a diffusing QW is presented.   


An outline of the paper is as follows: 
The quantization scheme of the SO(2) QW on integers is presented and its solution is derived. (Chapt. 2). The QW-MLE is developed and the analytic expression of the likelihood function is derived and graphed. Various relevant concepts (evolution loop, return probabilities, reluctant walk, translational symmetry etc) are introduced  (Chapt. 3). Complexity issues for the likelihood function estimation are presented in Chapt. 4, and a general discussion is given in Chapt. 5. Closing a two part appendix summarizes some mathematical properties needed.

\tableofcontents
\section{Quantization: from Classical to Quantum Random Walk}

Consider a classical random walk (CRW) on the lattice ${\mathbb Z}$ (or ${\mathbb Z}_N$) with 
${\mathcal H}_w \cong {\mathbb C}^{\mathbb Z}$ (or $\cong {\mathbb C}^N$ )\,
the walker's Hilbert space, and step and position operators $E_{\pm }$ and $L$
acting on the canonical basis respectively as $\left\vert
m\right\rangle \rightarrow E_{\pm }\left\vert m\right\rangle =\left\vert
m\pm 1\right\rangle ,$ and $\left\vert m\right\rangle \rightarrow
L\left\vert m\right\rangle =m\left\vert m\right\rangle $ 
(in the ${\mathbb Z}_N$ case the arithmetic is $(\mod N)$)\,.
Also the coin Hilbert space is taken to be $\mathcal{H}_c \cong {\mathbb Z}^{\{0,1\}}$
and the projection operators $P_{+}=\left\vert
0\right\rangle \left\langle 0\right\vert$\,, $P_{-}=\left\vert
1\right\rangle \left\langle 1\right\vert$\,, $~$ act in the
total space $\mathcal{H}_{c}\otimes \mathcal{H}_{w}\equiv \mathcal{H}_{cw}$
as $V_{cl}=P_{+}\otimes E_{+}+P_{-}\otimes E_{-}$ on coin-walker
density matrices $\rho _{c}\otimes \rho _{w}.$ The resulting CRW's $k$th step
 pmf 
$p_{cl}^{k}(m):= \left\langle m\right\vert \mathcal{E%
}_{cl}^{k}(\rho _{w})\left\vert m\right\rangle$ is
derived from the map $\rho
_{w}\rightarrow \mathcal{E}_{cl}^{k}(\rho _{w})=Tr_{c}[V_{cl}^{k}(\rho
_{c}\otimes \rho _{w})V_{cl}^{k\dag }].$

The so called ``$U$ quantization'' rule of the CRW modifies 
the classical evolution operator $V_{cl}$ by 
introducing the transformation \ $%
V_{cl}\rightarrow V_{q}=V_{cl}U\otimes \mathbb{I}_{w},$ where $U~\ $is the
unitary (reshuffling) matrix in coin space. The $k$'th step map reads $\rho
_{w}\rightarrow \mathcal{E}_{q}^{k}\left( \rho _{w}\right)
=Tr_{c}[V_{q}^{k}\left( \rho _{c}\otimes \rho _{w}\right) V_{q}^{k\dag }],$
\ and the related pmf is \smash{$p_{q}^{(k)}(m):=\left\langle m\right\vert \mathcal{E}%
_{q}^{k}(\rho _{w})\left\vert m\right\rangle .$} The quantum
effects in the resulting quantum random walk (QW) are attributed to the non-diagonal reshuffling matrix $U$\,, which
renders $V_{q}$ as a form of
entanglement creating operator (comparable to the Bell states generating
operator) \cite{ellqwentangle},\cite{ellinas and smyrnakis}.\\[1mm]

\subsection{The $\mathbb{Z}$-$SO(2)$ reluctant walk}
\label{sec:ZSO2walk}
If the coin matrix is chosen to be 
\[
U=%
\begin{pmatrix}
\cos \theta & \sin \theta \\ 
-\sin \theta & \cos \theta%
\end{pmatrix}
\]
with $U \in SO(2),$ and $\theta $ is angular parameter to be estimated, employing the
spectral properties of Euclidean algebra of generators i.e. $E_{\pm
}\left\vert \phi \right\rangle =e^{\pm i\widehat{\Phi }}\left\vert \phi
\right\rangle =e^{\pm i\phi }\left\vert \phi \right\rangle $\,, with
$\widehat{\Phi } = \left.\int\right._{0}^{2\pi }\left\vert \phi
\right\rangle \phi\left\langle \phi \right\vert \frac{d\phi }{2\pi }$\,,
etc. (see
Appendix C), the evolution operator $V_{q}=V_{cl}U\otimes \mathbb{I}_{w}$ reads 
\begin{equation}
V_{q}^{k}=\int\limits_{0}^{2\pi }\displaystyle{\left.\{ M(\phi ;\theta )^{k}\otimes \left\vert \phi
\right\rangle \left\langle \phi \right\vert \right.\}} \frac{d\phi }{2\pi } \label{one}
\end{equation}%
where 
\[ M(\phi ;\theta )=%
\begin{pmatrix}
e^{i\phi }\cos \theta & e^{-i\phi }\sin \theta \\ 
-e^{i\phi }\sin \theta & e^{-i\phi }\cos \theta%
\end{pmatrix}.
\]
The final result, in terms of $U_{n}(x),$ the Chebyshev polynomial of the
second kind of degree $n$\,, is summarized in\\[1mm]
\textbf{Proposition 1}: \textit{The }$%
\mathbb{Z}
$-$SO(2)$ \textit{QW's }$k$\textit{th step evolution unitary operator reads}%
\begin{equation*}
V_{q}^{k}=\left( 
\begin{array}{cc}
\mathcal{A}_{k} & \mathcal{B}_{k} \\ 
-\mathcal{B}_{k}^{\dagger } & \mathcal{A}_{k}^{\dagger }%
\end{array}%
\right) ,
\end{equation*}%
\textit{where} \ $\mathcal{A}_{k}(\widehat{\Phi };\theta )$, $\mathcal{B}%
_{k}(\widehat{\Phi };\theta ),$ \textit{with} \ $\mathcal{A}_{k}\mathcal{=}%
\cos \theta e^{i\widehat{\Phi }}\allowbreak U_{k-1}(\cos \theta \cos 
\widehat{\Phi })-U_{k-2}(\cos \theta \cos \widehat{\Phi })$\ \textit{and \ }$%
\mathcal{B}_{k}\mathcal{=}\sin \theta e^{-i\widehat{\Phi }}\allowbreak
U_{k-1}(\cos \theta \cos \widehat{\Phi });$ \textit{where \ $U_{n}(x)$ is the second
kind Chebyshev polynomial of order $n.$}
    The evolution map $\rho _{w}\rightarrow \mathcal{E}_{q}^{k}\left( \rho
    _{w}\right)
    =Tr_{c}[V_{q}^{k}\left( \rho
_{c}\otimes \rho _{w}\right) V_{q}^{k\dag }] $ for the initial coin state e.g. $\rho _{c}=\left\vert
    0\right\rangle \left\langle 0\right\vert ,$ reads 
    \begin{equation*}
    \mathcal{E}_{q}^{k}\left( \rho _{w}\right)=\mathcal{A}%
    _{k}\rho \mathcal{A}_{k}^{\dag }+\mathcal{B}_{k}\rho \mathcal{B}_{k}^{\dag }.
\end{equation*}
\hfill $\Box$ \\[1mm]

\textit{Proof}:\\
The characteristic equation of matrix $M,$ reads $P_{M}(\lambda )=\lambda
^{2}-2\xi \lambda +1=0,$ where $\xi =\frac{1}{2}Tr\left\{ M(\phi )\right\} .$
Utilizing the Cayley-Hamilton theorem \cite{bacry}, we obtain \ $%
P_{M}(M(\phi ))=M^{2}(\phi )-2\xi M(\phi )+\mathbb{I}=0,$ or $\ \ \ \
M^{2}(\phi )=2\xi M(\phi )-\mathbb{I}\mathbf{.}$

Last equation implies that any higher powers of $M(\phi )$ are linear
combinations of $M(\phi )$ and the identity matrix $\mathbb{I}$, i.e.

\begin{equation}
M^{k}(\phi )=M(\phi )U_{k-1}(\xi )-\mathbb{I}U_{k-2}(\xi ),  \label{3.1}
\end{equation}%
where $U_{k}(\xi )$ is a polynomial of degree $k$ in $\xi ,$ to be specified
shortly. Multiplying eq. (9) by $M(\phi )$ yields

\begin{eqnarray}
M^{k+1}(\phi ) &=&M^{2}(\phi _{1})U_{k-1}(\xi )-M(\phi )U_{k-2}(\xi )  \notag
\\
&=&[2\xi M(\phi )-\mathbb{I}]U_{k-1}(\xi )-M(\phi )U_{k-2}(\xi ),
\label{3.2}
\end{eqnarray}%
and from eq.$($\ref{3.1}) $\ $by substitution $k\rightarrow k+1,$ we obtain%
\begin{equation}
M^{k+1}(\phi )=M(\phi )U_{k}(\xi )-\mathbb{I}U_{k-1}(\xi ),  \label{3.3}
\end{equation}%
and by equating eqs.(\ref{3.2})(\ref{3.3}) \ and elaborating, we obtain

\begin{equation*}
M(\phi )[2\xi U_{k-1}(\xi )-U_{k-2}(\xi )-U_{k}(\xi )]=0,
\end{equation*}%
which, given that $M(\phi )\neq 0,$ yields the recurrence relation \ $2\xi
U_{k-1}(\xi )-U_{k-2}(\xi )-U_{k}(\xi )=0,$ which by substituting $%
k\rightarrow k+2,$ and multiplying by $\left( -1\right) ,$ becomes the
recurrence relation of Chebyshev polynomials,%
\begin{equation}
U_{k+2}(\xi )-2\xi U_{k+1}(\xi )+U_{k}(\xi )=0,\text{ }  \label{3.4}
\end{equation}

Then $k=1$ \ in eq.(\ref{3.1})$,$ leads to $2\xi M(\phi _{1})-\mathbb{I}%
=M(\phi _{1})U_{1}(\xi )-\mathbb{I}U_{0}(\xi ),$ which provides the
initialization $U_{0}(\xi )=1,$ \ $U_{1}(\xi )=2\xi .$ Summarizing $\left \{
U_{k}(\xi )\right \} _{k=0}^{\infty }~$are the second kind Chebyshev
polynomials, and by means of \ eq. (\ref{3.1}) would determine the density
matrix of QW at any step $k$ via equation (\ref{one}). (Note that eq. (1) for $k = 1$, is a form of half canonical decomposition of operator $V_q$\,, and that the walker's step operators are diagonal in the continuous basis
with delta-function orthogonality between kets (see Appendix C). This results in the $k$th power of $V_q$  involving the $k$th power in the kernel matrix $M(\phi)$).

By means of this result the power $k$th step evolution $\ \ $kernel matrix $%
M^{k}(\phi ;\theta )=\left( U(\theta )V_{cl}(\phi )\right) ^{k}$ \ is
obtained 
\begin{equation}
M^{k}(\phi ;\theta )=\left( 
\begin{array}{cc}
\cos \theta e^{i\phi }\allowbreak U_{k-1}(\xi )-U_{k-2}(\xi ) & \sin \theta
e^{-i\phi }\allowbreak U_{k-1}(\xi ) \\ 
-\sin \theta e^{i\phi }U_{k-1}(\xi ) & \cos \theta e^{-i\phi }U_{k-1}(\xi
)-U_{k-2}(\xi )%
\end{array}%
\right) .  \label{mmatrix}
\end{equation}

Referring to eq.(\ref{one}), and to the fact that matrix $M^{k}(\phi ;\theta
) $ depends only on variable $\phi $ we obtain via the spectral
decomposition of phase and step operators that the evolution
operator $V_{q}^{k} $ depends only on operator $\widehat{\Phi }$. More
explicitly $V_{q}^{k}$ depends only on the step operators $\widehat{E}_{\pm
}.$ Indeed in terms of the Euclidean algebra generators, the unitary
evolution operator reads $V_{q}^{k}$ 
\begin{equation}
V_{q}^{k}=\left( 
\begin{array}{cc}
\cos \theta e^{i\widehat{\Phi }}\allowbreak U_{k-1}(\cos \theta \cos 
\widehat{\Phi })-U_{k-2}(\cos \theta \cos \widehat{\Phi }) & \sin \theta
e^{-i\widehat{\Phi }}\allowbreak U_{k-1}(\cos \theta \cos \widehat{\Phi })
\\ 
-\sin \theta e^{i\widehat{\Phi }}U_{k-1}(\cos \theta \cos \widehat{\Phi }) & 
\cos \theta e^{-i\widehat{\Phi }}U_{k-1}(\cos \theta \cos \widehat{\Phi }%
)-U_{k-2}(\cos \theta \cos \widehat{\Phi })%
\end{array}%
\right) ,  \notag
\end{equation}%
or in terms of in terms of the step operators

\begin{eqnarray*}
&&V_{q}^{k}= \\
&&\left( 
\begin{array}{cc}
\begin{array}{c}
\widehat{E}_{+}\allowbreak U_{k-1}(\frac{1}{2}\cos \theta (\widehat{E}_{+}+%
\widehat{E}_{-}))\cos \theta \\ 
-U_{k-2}(\frac{1}{2}\cos \theta (\widehat{E}_{+}+\widehat{E}_{-}))%
\end{array}
& \widehat{E}_{-}\allowbreak U_{k-1}(\frac{1}{2}\cos \theta (\widehat{E}_{+}+%
\widehat{E}_{-}))\sin \theta \\ 
\begin{array}{c}
\\ 
-\widehat{E}_{+}U_{k-1}(\frac{1}{2}\cos \theta (\widehat{E}_{+}+\widehat{E}%
_{-}))\sin \theta%
\end{array}
& 
\begin{array}{c}
\widehat{E}_{-}U_{k-1}(\frac{1}{2}\cos \theta (\widehat{E}_{+}+\widehat{E}%
_{-}))\cos \theta \\ 
-U_{k-2}(\frac{1}{2}\cos \theta (\widehat{E}_{+}+\widehat{E}_{-}))%
\end{array}%
\end{array}%
\right) .
\end{eqnarray*}

More concisely \ 
\begin{equation*}
V_{q}^{k}=\left( 
\begin{array}{cc}
\mathcal{A}_{k} & \mathcal{B}_{k} \\ 
-\mathcal{B}_{k}^{\dagger } & \mathcal{A}_{k}^{\dagger }%
\end{array}%
\right) ,
\end{equation*}%
where \ $\mathcal{A}_{k}=\mathcal{A}_{k}(\widehat{\Phi };\theta )$, $%
\mathcal{B}_{k}=\mathcal{B}_{k}(\widehat{\Phi };\theta ),$ read explicitly 
\begin{eqnarray*}
\mathcal{A}_{k} &\mathcal{=}&\widehat{E}_{+}\allowbreak U_{k-1}(\frac{1}{2}%
\cos \theta (\widehat{E}_{+}+\widehat{E}_{-}))\cos \theta -U_{k-2}(\frac{1}{2%
}\cos \theta (\widehat{E}_{+}+\widehat{E}_{-})) \\
\mathcal{B}_{k} &\mathcal{=}&\widehat{E}_{-}\allowbreak U_{k-1}(\frac{1}{2}%
\cos \theta (\widehat{E}_{+}+\widehat{E}_{-}))\sin \theta \\
&&\text{or} \\
\mathcal{A}_{k} &\mathcal{=}&\cos \theta e^{i\widehat{\Phi }}\allowbreak
U_{k-1}(\cos \theta \cos \widehat{\Phi })-U_{k-2}(\cos \theta \cos \widehat{%
\Phi }) \\
\mathcal{B}_{k} &\mathcal{=}&\sin \theta e^{-i\widehat{\Phi }}\allowbreak
U_{k-1}(\cos \theta \cos \widehat{\Phi }).
\label{kraus}
\end{eqnarray*}

The evolution map $\rho _{w}\rightarrow \mathcal{E}_{q}^{k}\left( \rho
_{w}\right) $ for the initial coin state e.g. $\rho _{c}=\left\vert
0\right\rangle \left\langle 0\right\vert ,$ reads 
\begin{equation*}
\mathcal{E}_{q}^{k}\left( \rho _{w}\right) =Tr_{c}[V_{q}^{k}\left( \rho
_{c}\otimes \rho _{w}\right) V_{q}^{k\dag }]=Tr_{c}[V_{q}^{k}\left( 
\begin{array}{cc}
\rho _{w} & 0 \\ 
0 & 0%
\end{array}%
\right) V_{q}^{k\dag }],
\end{equation*}%
and is provided by means of the positive trace preserving map 
\begin{equation*}
\rho _{w}\rightarrow \mathcal{E}_{q}^{k}\left( \rho _{w}\right) =\mathcal{A}%
_{k}\rho \mathcal{A}_{k}^{\dag }+\mathcal{B}_{k}\rho \mathcal{B}_{k}^{\dag }, \label{channel}
\end{equation*}%
where the generators ($\mathcal{A}_{k},\mathcal{B}_{k})$ of the map are
normal operators (commute with their Hermitian conjugate), and satisfy the
trace preserving relation $\mathcal{A}_{k}^{\dag }\mathcal{A}_{k}+\mathcal{B}%
_{k}^{\dag }\mathcal{B}_{k}=\mathbb{I}_{w}.$ \hfill $\Box$\\

The occupation probabilities, expressed in terms of the parameter $\lambda
(\theta )=\cos \theta ,$ as an argument of the ${}_2F_1$ hypergeometric function,
which for the parameter ranges
applicable is truncated to a polynomial in $\lambda$\,, \smash{$Y_{d}^{(2k)}(\lambda )$}\,, are exactly determined in\\[1mm]
\noindent
\textbf{Proposition 2}\textit{: The pmf }$p^{(k)}(d;\lambda )$\textit{\
assigns zero probability to odd steps, and even steps }$k$ \textit{have equal
occupation probabilities for sites distanced \ }$\pm |d|$\textit{\ units away
from the initial }$0$\textit{\ site}: 
\begin{equation}
p^{(k)}(d;\lambda )=(Y_{d}^{(k)}(\lambda ))^{2}+(Y_{d-1}^{(k-1)}(\lambda
))^{2}-2Y_{d}^{(k)}(\lambda )Y_{d-1}^{(k-1)}(\lambda )\cos \theta
\label{mainpdf}    
\end{equation}%
\textit{where \ the polynomial }$Y_{d}^{(k)}$ \,($= Y_{-d}^{(k)}$\,) \textit{reads}%
\begin{equation*}
Y_{d}^{(k)}(\lambda )=\lambda ^{k}\tbinom{2k}{\frac{2k+d}{2}}\text{\textit{\ 
}}_{2}F_{1}(\frac{d-2k}{2},\frac{-d-2k}{2},-2k,{\lambda ^{-2}}).
\end{equation*}
\mbox{}\\[-3mm]\mbox{}\hfill $\Box$

\noindent
\textit{Proof}:\\
Explicit evaluation of pmf: we proceed with the evaluation of $%
p^{(k)}(x|\theta )$ which 
reads,%
\begin{eqnarray*}
p^{(k)}(x|\theta ) &=&\int\limits_{0}^{2\pi }\int\limits_{0}^{2\pi
}e^{id(\phi _{1}-\phi _{2})}U_{k-1}(\xi _{1}){}U_{k-1}(\xi _{2}){}Tr_{c} 
\left[ M(\phi _{1};\theta )\rho _{c}M(\phi _{2};\theta )^{\dagger }\right] 
\frac{d\phi _{1}d\phi _{2}}{\left( 2\pi \right) ^{2}} \\
&&-\int\limits_{0}^{2\pi }\int\limits_{0}^{2\pi }e^{id(\phi _{1}-\phi
_{2})}U_{k-1}(\xi _{1}){}U_{k-2}(\xi _{2}){}Tr_{c}\left[ M(\phi _{1};\theta
)\rho _{c}\right] \frac{d\phi _{1}d\phi _{2}}{\left( 2\pi \right) ^{2}} \\
&&-\int\limits_{0}^{2\pi }\int\limits_{0}^{2\pi }e^{id(\phi _{1}-\phi
_{2})}U_{k-2}(\xi _{1}){}U_{k-1}(\xi _{2}){}Tr_{c}\left[ \rho _{c}M(\phi
_{2};\theta )^{\dagger }\right] \frac{d\phi _{1}d\phi _{2}}{\left( 2\pi
\right) ^{2}} \\
&&+\int\limits_{0}^{2\pi }\int\limits_{0}^{2\pi }e^{id(\phi _{1}-\phi
_{2})}U_{k-2}(\xi _{1}){}U_{k-2}(\xi _{2}){}\frac{d\phi _{1}d\phi _{2}}{%
\left( 2\pi \right) ^{2}}.
\end{eqnarray*}

To proceed we need to evaluate the three traces of the square bracketed
expressions above.\ \ For the infinite case, recalling the matrix $M$ from
eq. (\ref{mmatrix}) and choosing the initial coin state to be $\left \vert
c\right \rangle =\left \vert 0\right \rangle ,$ we get 
\begin{eqnarray*}
Tr_{c}\left[ M(\phi _{1};\theta )\rho _{c}M(\phi _{2};\theta )^{\dagger }%
\right] &=&e^{i\left( \phi _{1}-\phi _{2}\right) }, \\
Tr_{c}\left[ M(\phi _{1};\theta )\rho _{c}\right] &=&\cos \theta e^{i\phi
_{1}}, \\
Tr_{c}\left[ \rho _{c}M(\phi _{2};\theta )^{\dagger }\right] &=&\cos \theta
e^{-i\phi _{2}}.
\end{eqnarray*}

This results into the next expression for the distribution function, in
terms of some factorized integrals,

\begin{align*}
p^{(k)}(x|\theta )=& \,\left\vert \int_{0}^{2\pi }e^{i{d}\phi
}U_{k\!-\!1}(\lambda \cos \phi )\frac{{d}\phi }{2\pi }\right\vert
^{2}+\left\vert \int_{0}^{2\pi }e^{i{d}\phi }U_{k\!-\!2}(\lambda \cos \phi )%
\frac{{d}\phi }{2\pi }\right\vert ^{2}- \\
& \,-2\lambda  \left\{ \int_{0}^{2\pi }e^{i(d\!+\!1)\phi
_{1}}U_{k\!-\!1}(\lambda \cos \phi _{1})\frac{{d}\phi _{1}}{2\pi }\cdot
\int_{0}^{2\pi }e^{-i{d}\phi _{2}}U_{k\!-\!2}(\lambda \cos \phi _{2})\frac{{d%
}\phi _{2}}{2\pi }\right \},
\end{align*}%

where $\lambda =\cos \theta $. Because of independence of the sign of $d$ in
the exponential, the indicated integrals above are real e.g.: 
\begin{equation*}
\int_{0}^{2\pi }U_{k}(\lambda \cos \phi )e^{+id\phi }\frac{{d}\phi }{2\pi }%
\,\equiv \int_{0}^{2\pi }U_{k}(\lambda \cos \phi )\cos (d\phi )\frac{{d}\phi 
}{2\pi }=Y_{|d|}^{(k)}(\lambda ),
\end{equation*}%
leading to the analytic form of the pmf 
\begin{equation*}
p^{(k)}(x|\theta )=\,(1-\lambda ^{2})\left( Y_{|d-1|}^{(k-1)}(\lambda
)\right) ^{2}+\left( Y_{|d|}^{(k-2)}(\lambda )-\lambda
Y_{|d+1|}^{(k-1)}(\lambda )\right) ^{2}\,.
\end{equation*}

The probability distribution of the walk is solved for analytically and is
evaluated in terms of the functions

\begin{equation}
Y_{d}^{(k)}(\lambda )=\int_{0}^{2\pi }U_{k}(\lambda \cos \phi )\cos (d\phi )%
\frac{d\phi }{2\pi }.  \label{y}
\end{equation}

This integral can be evaluated by writing the $k$th order Chebyshev
polynomial of the second kind in a series form, expanding $\cos (d\phi )$ in
powers of $\cos \phi $ and then performing the integral on the remaining
terms involving $\phi $. Doing this gives us a new definition for $%
Y_{d}^{(k)}(\lambda )$ in terms of a series

\begin{equation*}
Y_{d}^{(k)}(\lambda )=\frac{d}{2}\sum_{p=0}^{\left\lfloor \frac{k}{2}%
\right\rfloor }\sum_{q=0}^{\left\lfloor \frac{k}{2}\right\rfloor }\frac{%
(-1)^{p+q}}{d-q}\binom{k-p}{p}\binom{d-q}{q}\binom{k+d-2p-2q}{\frac{1}{2}%
(k+d)-p-q}\lambda ^{k-2p}.
\end{equation*}

This double sum series is not very useful. To simplify it recall that the $k 
$th order Chebyshev polynomial of the second kind of an arbitrary argument $%
x $ can be written as a terminating series with the following form%
\begin{equation*}
U_{k}(x)=\sum_{n=0}^{\left\lfloor \frac{k}{2}\right\rfloor }(-1)^{n}\binom{%
k-n}{n}(2x)^{k-2n}.
\end{equation*}

Therefore we can rewrite the integrand of eq.\ref{y} as%
\begin{equation*}
Y_{d}^{(k)}(\lambda )=\int_{0}^{2\pi }\left( \sum_{n=0}^{\left\lfloor \frac{k%
}{2}\right\rfloor }(-1)^{n}\binom{k-n}{n}(2\lambda \cos \phi )^{k-2n}\right)
\cos (d\phi )\frac{d\phi }{2\pi }
\end{equation*}

Since this is a terminating sum we can freely exchange the order of the
integration and summation yielding 
\begin{equation}
Y_{d}^{(k)}(\lambda )=\sum_{n=0}^{\left\lfloor \frac{k}{2}\right\rfloor
}(-1)^{n}\binom{k-n}{n}(2\lambda )^{k-2n}\left( \int_{0}^{2\pi }(\cos \phi
)^{k-2n}\text{ }\cos (d\phi )\frac{d\phi }{2\pi }\right) .  \label{yy}
\end{equation}

We will convert the above real integral into a complex contour integral and
use the calculus of residues to evaluate it. To this end let the integral 
\begin{equation*}
I=\int_{0}^{2\pi }(\cos \phi )^{k-2n}\text{ }\cos (d\phi )\frac{d\phi }{2\pi 
}
\end{equation*}%
and make the substitutions $z=e^{i\phi },$ $\frac{dz}{d\phi }=iz$ \ and $%
\cos \phi =\frac{z+z^{-1}}{2},$ $\cos (d\phi )=\frac{z^{d}+z^{-d}}{2}$ \
transforming the integral into

\begin{equation*}
I=2^{-p-1}\frac{1}{2\pi i}\int\limits_{\Gamma }z^{p}\left( 1+\frac{1}{z^{2}}%
\right) ^{p}\left( z^{d}+z^{-d}\right) z^{-1}dz,
\end{equation*}%
where $p=k-2n$ (note $p\geq 0).$ The integrand only contains polynomial
terms and so we can clearly see that it is regular everywhere except for at
the origin where there is a pole. To find the residue of this pole we start
by using a binomial series to expand the term raised to the power $p,$ i.e.

\begin{equation*}
\left( 1+\frac{1}{z^{2}}\right) ^{p}=\sum_{j=0}^{p}\binom{p}{j}z^{-2j}.
\end{equation*}

This then splits the integrand into two series \ 
\begin{equation}
z^{p}\left( 1+\frac{1}{z^{2}}\right) ^{p}\left( z^{d}+z^{-d}\right)
z^{-1}=\sum_{j=0}^{p}\binom{p}{j}z^{p+d-2j-1}+\sum_{j=0}^{p}\binom{p}{j}%
z^{p-d-2j-1}.  \label{pole}
\end{equation}

It should be noted that since the only change in the exponent of each series
is the sign of $d$, the actual sign of $d$ does not matter. Therefore,
without loss of generality, we can take $d\geq 0$. We now look for the
residue by examining the coefficient of the $z^{-1}$ term. First we must
consider when this term actually exists in the series (a non-zero residue).
Since $-2j-1$ is odd, it is necessary that $p\equiv d$ $\pmod 2$. Since $%
p=k-2n$ and $-2n\equiv 0$ $\pmod 2$ this means that $k\equiv d$ $\pmod 2$
(i.e. $d$ and $k$ have to be both even or both odd for a non-zero residue, a
fact we already knew). Now, taking $p\equiv d$ $\pmod 2$, over the range of
the summation, $p+d-2j-1$ and $p-d-2j-1$ form monotone decreasing series
taking the value of every odd integer in the interval between when $j=0$ and 
$j=p$. We want $-1$ to lie in this interval so we get four conditions on $p$
and $d$%
\begin{eqnarray*}
p+d &\geq &0\geq d-p \\
p-d &\geq &0\geq -d-p,
\end{eqnarray*}%
which are reduced to only one $p\geq d$ which is satisfied and implies that $%
n>\frac{k-d}{2}$. Now we look at what the actual value of the residue at the
origin is, c.f. eq.(\ref{pole}), and find $\ $Res($z=0)=\binom{p}{\frac{1}{2}%
(p+d)}.$

So we can now evaluate the integral to be 
\begin{equation*}
I=2^{-p}\binom{p}{\frac{1}{2}(p+d)}
\end{equation*}%
Returning to eq.(\ref{yy}) we obtain

\begin{equation}
Y_{d}^{(k)}(\lambda )=\sum_{n=0}^{\frac{k-d}{2}}(-1)^{n}\binom{k-n}{n}\binom{%
k-2d}{\frac{1}{2}(k+d)-n}\lambda ^{k-2n}.  \label{yfinal}
\end{equation}

As a final remark note that in eq. (17) the term with the lowest power
of $\lambda $ occurs for the greatest value of $n$ which is $\lambda ^{k-2%
\frac{k-d}{2}}=\lambda ^{d},$ i.e. the polynomial $Y_{d}^{(k)}$ has a
monomial factor $d$ that factorizes out of the entire series.

Next we show \ how $Y_{d}^{(k)}$ \ in the single sum form as above is
expressed in terms of the Gaussian hypergeometric series

\begin{equation*}
_{2}F_{1}(a,b;c;z)=\sum_{n=0}^{\infty }\frac{(a)_{n}(b)_{n}}{(c)_{n}}\frac{%
z^{n}}{n!}.
\end{equation*}

Note first that this series terminates if one of its first two arguments, a
or b, is a negative integer. If e.g. $a=-m$ \ then the infinite series
collapses to

\begin{equation*}
_{2}F_{1}(a,b;c;z)=\sum_{n=0}^{m}(-1)^{m}\binom{m}{n}\frac{(b)_{n}}{(c)_{n}}%
\frac{z^{n}}{n!}.
\end{equation*}

In the process of putting eq.(\ref{yfinal}) into the form of the polynomial
above we will make use of the following identities for Pocchammer symbols 
\begin{eqnarray*}
a! &=&(-1)^{a}(-a)_{a} \\
(a)_{m} &=&(a)_{l}(a+l)_{m-l} \\
\binom{m}{n} &=&\frac{(-m)_{n}}{(-n)_{n}}.
\end{eqnarray*}

This leads to the following form of $Y_{d}^{(k)}$ 
\begin{equation*}
Y_{d}^{(k)}(\lambda )=\lambda ^{k}\sum_{n=0}^{\frac{k-d}{2}}(-1)^{n}\binom{-%
\frac{d-k}{2}}{n}\frac{\left( \frac{-d-k}{2}\right) _{n}}{\left( -k\right)
_{n}}\lambda ^{-2n},
\end{equation*}%
which is identified with the form 
\begin{equation*}
Y_{d}^{(k)}(\lambda )=\lambda ^{k}\binom{k}{\frac{k+d}{2}}\text{ }_{2}F_{1}(%
\frac{d-k}{2},\frac{-d-k}{2};k;\frac{1}{\lambda ^{2}})
\end{equation*}%
as required.\hfill $\square $\\

\noindent
\textbf{Remark 1}\\
By means of the notations $(Y_{d}^{(k)}(\lambda ))^{2}=p_{d}^{(k)}$and $\
(Y_{d-1}^{(k-1)}(\lambda ))^{2}=p_{d}^{(k-1)},$ the pmf in eq.(\ref{mainpdf}%
), reads 
\begin{equation*}
p^{(k)}(d|\lambda )=p_{d}^{(k)}+p_{d}^{(k-1)}-2\sqrt{p_{d}^{(k)}p_{d}^{(k-1)}%
}\cos \theta =\left\vert p_{d}^{(k)}-p_{d}^{(k-1)}e^{i\theta }\right\vert
^{2}.
\end{equation*}%
This is identified with the law of cosines $\gamma ^{2}=\alpha ^{2}+\beta
^{2}-2\alpha \beta \cos \theta =\left\vert \alpha ^{2}-\beta ^{2}e^{i\theta
}\right\vert ^{2},$ for a triangle of sides $\alpha ,\beta ,\gamma $ of
respective lengths $Y_{d}^{(k)},Y_{d-1}^{(k-1)},\sqrt{p_{d}^{(k)}}$ and of
angle $\theta $ between sides $\alpha $ and $\beta .$ The values $0,\frac{%
\pi }{2},\pi ,\frac{3\pi }{2},\theta $ of the reshuffling angle $\theta $
and their cosines $\lambda (\theta )=1,0,-1,0,\cos \theta ,$ correspond to
QWs with diagonal, off-diagonal coin matrices $\mathbb{I},Y=i\sigma _{y},-%
\mathbb{I},-Y,R(\theta );$ (for the effect of non-diagonality of reshuffling
matrix to the quantization of a CRW in relation to quantization rules see, 
\cite{ellinas and smyrnakis}). The respective probabilities \ distributions $%
p_{d}^{(k)}\equiv \gamma ^{2}=(\alpha -\beta )^{2}$ $[CRW],\alpha ^{2}+\beta
^{2}$ $[QW],(\alpha +\beta )^{2}$ $[CRW],\alpha ^{2}+\beta ^{2}$ $[QW]$ and $%
\alpha ^{2}+\beta ^{2}-2\alpha \beta \cos \theta $ $[QW],$ are associated,
as indicated, to CRW ($\gamma :$ exact squares) or QW ($\gamma :$ not exact
squares). The conditional appearance of law of cosines in the final
expression of pmf is a manifestation of the quantum character of the walk.\\

\noindent
\textbf{Remark 2}\\
By means of the last final relation and the contiguity relations of the
hypergeometric function we can cast the final expression for the pmf of the
reluctant QW as follows

\begin{equation*}
p_{2d}^{(2k)}(\lambda )=\left( \lambda ^{2k}\binom{2k}{d}\right) ^{2}\times
\left( _{2}F_{1}(d-k,-d-k;-2k;\frac{1}{\lambda ^{2}})^{2}+\lambda ^{2}\text{ 
}_{2}F_{1}(d-k-1,-d-k;-2k;\frac{1}{\lambda ^{2}})^{2}\right) .
\end{equation*}

Propositions 1 and 2 provide an exact analytic solution to the pmf of the $%
\mathbb{Z}
$-$SO(2)$ QW for arbitrary parameter $\theta $  and any displacement $d$ away
from the origin. \ The novel feature is the
maximum of the pmf achieved at the center of the coordinates i.e. $\lambda
=r=0$\,, and in general, for given $\lambda$, also at a specific $d$
determined by the ratio $r:= d/k$\, which we term the reluctance.
This ``reluctant walker'' behaviour will allow for the parameter estimation of $\theta $ (see below).
\mbox{}\\[-.7cm]

\begin{center}
\includegraphics[width=8cm]{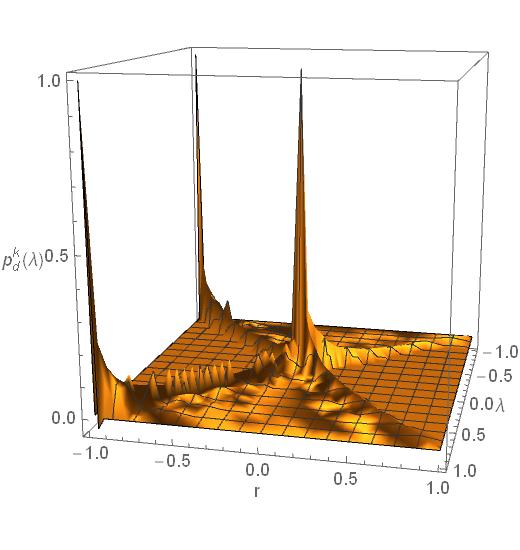}
\begin{quotation}
\noindent
Fig. 1 Smoothed surface plot for fixed  $k=100$. Points plotted are
coordinates and height values $\{\lambda ,r=d/100,p^{k=100}(d;\lambda \}\}$\,.
\end{quotation}
\includegraphics[width=6cm]{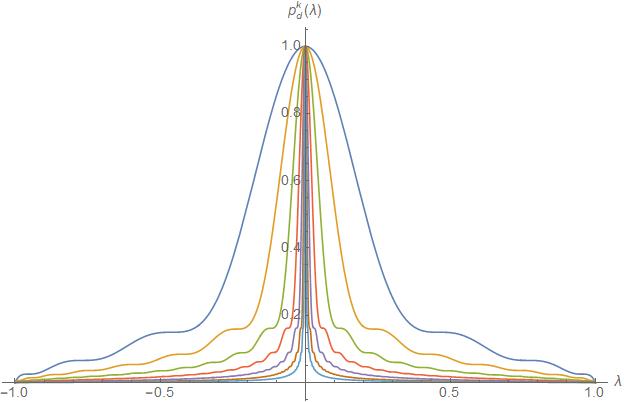}\hskip.5cm
\includegraphics[width=6cm]{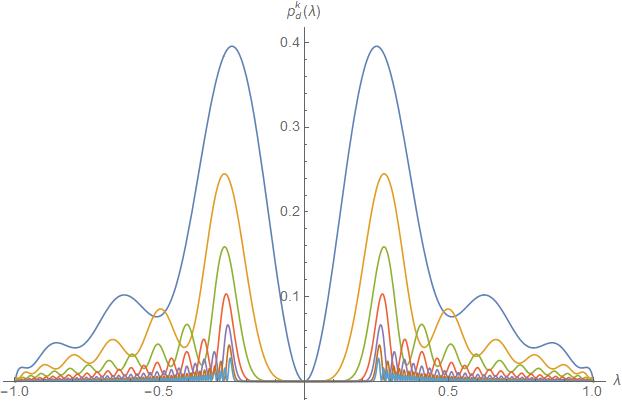}
\begin{quotation}
{\noindent
Fig. 2 Upper panel: plots of the pmf with respect to $\lambda $
 for $d=0$. The curves have $k$ values of $k=2^{\ell}$,$%
\ell=3,\ldots ,9$\,.\\ \noindent
Lower panel: same as above, for fixed 
 $r=d/k=0.25$ }
\end{quotation}

\end{center}
 The complicated form of the pmf issued in equation (\ref{mainpdf}), enforces numerical plotting of the maximization of the likelihood
function (see figures), instead of an analytic calculation of the
maximum. This inconvenience stems from the fact that the pmf does not
belong to the exponential family of probability density functions, for which
the MLE usually admits an analytic treatment \cite{roussas}.

\section{Operational approach to the MLE\ problem}

Turning to the MLE of $\theta $ parameter for the
group theoretical models of QW above leads to the following considerations.

\subsection{MLE: $\mathbb{Z}$-$SO(2)$ case.}
Let the position
projection operator $L^{(d)}=\left\vert s+d\right\rangle \left\langle
s+d\right\vert =E_{+}^{s}\left\vert d\right\rangle \left\langle d\right\vert
E_{+}^{s\dag }=Ad(E_{+}^{s})\left\vert d\right\rangle \left\langle
d\right\vert ,$ which would provide the probability that after $k$ steps the
walker reluctantly moved $d$ steps away from its initial cite.

The CP evolution map $\mathcal{E}_{q}^{k}$ is generated by step operators $%
E_{\pm }$ satisfying the translation invariant condition $Ad(E_{+}^{\dag
s})\circ \mathcal{E}_{q}^{k}=\mathcal{E}_{q}^{k}\circ (Ad(E_{+}^{\dag s})),$
so the occupation probability for initial state $\left\vert \Psi
\right\rangle =\left\vert 0\right\rangle ,$ reads 
\begin{equation*}
p^{(k)}(d;\theta )=Tr[L^{(d)}(\rho _{w}^{(k)})]=\left\langle d\right\vert 
\mathcal{E}_{q}^{k}(\left\vert 0\right\rangle \left\langle 0\right\vert
)\left\vert d\right\rangle .
\end{equation*}%
The likelihood function defined for a set of position points denoted by
vector $\overrightarrow{x}=\left( x_{1},...,x_{n}\right) ,$ reads 
\[
\mathcal{L}_{n}^{(k)}(\theta |x)\equiv \prod\limits_{j=1}^{n}p^{(k)}(x_{j}|\theta );
\]
where we denote $p^{(k)}(x_{j};\theta )$ \ by $p^{(k)}(x_{j}|\theta )$ and
designate the likelihood function as $\mathcal{L}_{n}^{(k)}(\theta |x)$. Next let
positions $\overrightarrow{x}$ be occupied by walker after $k$ steps.
Maximization of the logarithm of likelihood $l_{n}^{(k)}(\theta ;x)=\log
L_{n}^{(k)}(\theta ),$ requires the solution of equation $\frac{\partial }{%
\partial \theta }l_{n}^{(k)}(\theta )=0$, for each $j$ and the determination
of the $\theta $ roots of function $p^{(k)}(x_{j}|\theta ).$

Let us choose an \textit{evolution loop} for the QW i.e. $L^{(k)}\mathcal{=}%
\rho _{w}^{(0)}=|x\rangle \langle x|,$ where $|x\rangle ,$ $x\in 
\mathbb{Z}
,$ a basis vector. This choice implies that initial and final state for QWer
\ be the same, say some $|x\rangle ,$ $x\in 
\mathbb{Z}
.$ This choice of loop 
implies a drastic simplification for the likelihood
function since now the lattice position vector becomes $\ \overrightarrow{x}%
=\left( x,...,x\right) $ and imposes the simplification 
\begin{equation}
l_{n}^{(k)}\left( \theta |x\right) =\log \left( p^{(k)}\left( x|\theta
\right) \right) ^{n}=n\log p^{(k)}(x|\theta ).  \label{nk}
\end{equation}%
Since $p^{(k)}\in \left[ 0,1\right] ,$ we obtain $l_{n}^{(k)}=\log
p^{(k)}<0, $ thus the likelihood must be negative and its maximization
requires $\left( l_{n}^{(k)}\left( \theta |x\right) \right) ^{\prime \prime
}<0,$ where the derivatives with respect to $\theta $ are denoted by primes.\\

The evolution loop idea implies the involvement of
return probabilities in the evaluation of likelihood function of the
reluctant QW resulting into a factorization of $l_{n}^{(k)}$
 into
factors depending separately on $n$  and $k,$ as in equation (%
\ref{nk}). Likelihood maximization amounts to maximization of the distribution %
$p^{(k)}(x|\theta )$,  for given $k\in 
\mathbb{N}$
for those $\theta \in \lbrack -\pi ,\pi )$  for which
the distribution satisfies the positivity inequality%
\begin{equation}
p^{(k)\prime 2}-p^{(k)\prime \prime }p^{(k)}>0.
\label{firstinequality}
\end{equation}

 The complicated form of the pmf issued in equation (\ref{mainpdf}), enforces numerical plotting of the maximization of the likelihood
 function (see figures), instead of an analytic calculation of the
 maximum. This inconvenience stems from the fact that the pmf does not
belong to the exponential family of probability density functions, for which
 the MLE usually admits an analytic treatment \cite{roussas}.  

 Finally we point out an alternative experimental scenario for likelihood
estimation, which does not require the use of translational invariance (to
be explained below) in constructing probabilities for multiple QW trials
sampled such that there is net zero displacement at various positions.
Rather, in each trial the QW is allowed to run from the origin $x=0$, and
then a measurement is taken after $k$ steps to establish whether or not it
has returned. In this scenario, with $p:=p^{(k)}(0,\lambda )$ and $n_{0}$
recordings of measurements at $x=0$ out of $n$ trials, the log likelihood is

\begin{equation}
l=n_{0}\log p+(n-n_{0})\log (1-p)
\label{logl}
\end{equation}
so that, at the optimum ( assuming now $p^{\prime }\ne0$) , the odds $p/(1-p)$
of return are the ratio $n_{0}:(n-n_{0})$; that is, we simply have in terms
of the relative frequency of return $f:=n_{0}/n$,%
\begin{equation}
p^{(k)}(0,\lambda )=f.
\label{f}
\end{equation}

The possible solutions in $\lambda =\cos \theta $ are therefore the level
set at height $f$ of the polynomial function $p$ (Proposition $2$ and
equation (\ref{mainpdf}), and provided $-l^{\prime \prime }>0$). Such a ``high
reluctance" protocol might be appropriate for small parameter values, where
the first intersection is provided by the dominant peak of the pmf around
the origin $\lambda =0$ (see figure 2(a)).

\subsection{QW-MLE with return probabilities}
The treatment of this subsection is presented in order to show transparently the ideas of return probabilities, of closed loops and of the symmetry of translation invariance of the QW and how all these are used for constructing likelihood function of MLE task. 

Start with the QW density matrix at $k$-th step $\rho _{w}\rightarrow \mathcal{E}%
_{q}^{k}\left( \rho _{w}\right) $ that reads,

\[
\mathcal{E}_{q}^{(k)}\left( \rho _{w}\right) =Tr_{c}[V_{q}^{k}\left( \rho
_{c}\otimes \rho _{w}\right) V_{q}^{k\dag }]. 
\]

Let the generic variables $a,b\in 
\mathbb{Z}
,$ \ and define the projectors $P_{a}=|\left. a\right\rangle \left\langle
a\right. |$ \ and $P_{b}=|\left. b\right\rangle \left\langle b\right. |.$ If
initially $\rho _{w}=|\left. a\right\rangle \left\langle a\right. |$ then
the transition probability from site $a$ to site $b$ after $k\mathcal{\ }$
steps is obtained by means of a projection measurement as

\begin{eqnarray*}
p^{(k)}\left( a,b\text{ }|\theta \right) &=&Tr[P_{b}\mathcal{E}%
_{q}^{(k)}(|\left. a\right\rangle \left\langle a\right. |)] \\
&=&\left\langle b\right\vert \mathcal{E}_{q}^{(k)}(|\left. a\right\rangle
\left\langle a\right. |)\left\vert b\right\rangle 
_{q}^{(k)})_{a,b}.
\end{eqnarray*}%

Then we proceed by noting that the position states are generated from the
zero cite (vacuum) state as $|\left. a\right\rangle =e^{i\widehat{\Phi }%
a}|\left. 0\right\rangle ,$ $|\left. b\right\rangle =e^{i\widehat{\Phi }%
b}|\left. 0\right\rangle $ \ or in terms of the step operators $|\left.
a\right\rangle =E_{+}^{a}|\left. 0\right\rangle $ and $|\left.
b\right\rangle =E_{+}^{b}|\left. 0\right\rangle ,$ it follows that

\begin{eqnarray*}
|\left. a\right\rangle \left\langle a\right. | &=&E_{+}^{a}|\left.
0\right\rangle \left\langle 0\right. |E_{+}^{a\dagger
}=E_{+}^{a}P_{0}E_{+}^{a\dagger },\  \\
\text{and }|\left. b\right\rangle \left\langle b\right. |
&=&E_{+}^{b}|\left. 0\right\rangle \left\langle 0\right. |E_{+}^{b\dagger
}=E_{+}^{b}P_{0}E_{+}^{b\dagger }.
\end{eqnarray*}%
\ As shown in Prop. 1, the evolution channel is 
\[
\mathcal{E}_{q}^{(k)}(\rho _{w})=\mathcal{A}_{k}(\widehat{\Phi };\theta
)\rho _{w}\mathcal{A}_{k}(\widehat{\Phi };\theta )^{\dagger }+\mathcal{B}%
_{k}(\widehat{\Phi };\theta )\rho _{w}\mathcal{B}_{k}(\widehat{\Phi };\theta
)^{\dagger }. 
\]%
Note that the generators $\mathcal{A}_{k}(\widehat{\Phi };\theta )$, $%
\mathcal{B}_{k}(\widehat{\Phi };\theta ),$ are commuting with step operators 
$E_{\pm }$ \ i.e. $E_{+}\mathcal{A}_{k}(\widehat{\Phi };\theta )=E_{+}%
\mathcal{A}_{k}(\widehat{\Phi };\theta )$, and $E_{+}\mathcal{B}_{k}(%
\widehat{\Phi };\theta )=E_{+}\mathcal{B}_{k}(\widehat{\Phi };\theta ),$ a
property that give rise to translation invariance.

Then we have 
\begin{eqnarray*}
p^{(k)}\left( a,b|\theta \right) &=&Tr\left( E_{+}^{b}P_{0}E_{+}^{b\dagger }%
\mathcal{E}_{q}^{k}(E_{+}^{a}P_{0}E_{+}^{a\dagger }\right) \\
&=&Tr\left( E_{+}^{b}P_{0}E_{+}^{b\dagger }E_{+}^{a}\mathcal{E}%
_{q}^{k}(P_{0}\right) E_{+}^{a\dagger } \\
&=&Tr\left( \left( E_{+}^{a\dagger }E_{+}^{a}\right) P_{0}\left(
E_{+}^{b\dagger }E_{+}^{b}\right) ^{\dagger }\mathcal{E}_{q}^{(k)}(P_{0})%
\right) ,
\end{eqnarray*}%
where the second equation above follows again from mentioned commutativity.

Elaborating last equation is cast in the form:

\[
p^{(k)}\left( a,b|\theta \right) =Tr\left( e^{-i\widehat{\Phi }%
(a-b)}P_{0}e^{i\widehat{\Phi }(a-b)}\mathcal{E}_{q}^{(k)}(P_{0})\right)
\]

Note that if $a-b=d\in 
\mathbb{Z}
$ , i.e. if we have a path of $d$ displacement units then $
p^{(k)}\left( b,b+d|\theta \right) \equiv
p_{d}^{(k)}(\theta )$ 
If further $d=0$ we have a return probability \ $%
p^{(k)}\left( a,a|\theta \right) =p_{0}^{(k)}(\theta )=Tr(P_{a}\mathcal{E}%
_{q}^{(k)}(P_{a})).$  \ 

The likelihood function employing transition probabilities as above for
paths with initial site $x_{in}^{j}$ \ \ and final site  $%
x_{f}^{j}=x_{in}^{j}+d_{j},$ i.e. a site displaced by $d_{j}$ integers , reads%
\begin{equation}
\mathcal{L}_{n}^{(k)}(\theta |\left\{ d_{1},...,d_{n}\right\} )\equiv
\prod\limits_{j=1}^{n}p^{(k)}(d_{j}|\theta ).
\end{equation}

Assuming that the statistical model $\rho _{w}\rightarrow \mathcal{E}%
_{q}^{(k)}[\theta ](\rho _{w})\equiv \rho _{w}^{(k)}(\theta )$ employed for
QW-MLE contains the data generation statistical assumption that $d_{j}=d\in 
\mathbb{Z}
,$  implies that the likelihood function is computed in terms of transition 
probabilities of paths of equal length $d$. These probabilities are computed by standard projection
operators $P_{x_{in}^{j}}=|\left. x_{in}^{j}\right\rangle \left\langle
x_{in}^{j}\right. |$ \ and $P_{x_{f}^{j}}=|\left. x_{f}^{j}\right\rangle
\left\langle x_{f}^{j}\right. |$ that perform quantum measurements on the
initial and the $k$-th step walker density matrix. \ In such a case the
likelihood function is cast in the form%
\begin{equation}
\mathcal{L}_{n}^{(k)}(\theta |\left\{ d_{1}=d,...,d_{n}=d\right\} )=\left(
p^{(k)}(d|\theta )\right) ^{n}.
\end{equation}

For the special case of the return probability with $x_{in}^{j}=x_{f}^{j}=x$ the likelihood function should be computed at the point $d=0.$

Before closing this section we show how to compute  the distribution of return probabilities formally in terms of the
Kraus generators of the evolution channel. The return probability of zero displacement reads for some integer $x$ as follows:

\begin{eqnarray*}
p^{(k)}(x,x|\theta ) &=&Tr(|\left. x\right\rangle \left\langle
x\right. |\mathcal{E}_{q}^{(k)}(|\left. x\right\rangle \left\langle x\right.
|))=\left\langle x\right. |\mathcal{E}_{q}^{(k)}(|\left. x\right\rangle
\left\langle x\right. |)|\left. x\right\rangle \\
&=&\left\langle x\right. |\left( \mathcal{A}_{k}|\left. x\right\rangle
\left\langle x\right. |\mathcal{A}_{k}^{\dagger }+\mathcal{B}_{k}|\left.
x\right\rangle \left\langle x\right. |\mathcal{B}_{k}^{\dagger }\right)
|\left. x\right\rangle \\
&=&\left\langle x\right. |\mathcal{A}_{k}|\left. x\right\rangle \left\langle
x\right. |\mathcal{A}_{k}^{\ast }|\left. x\right\rangle +\left\langle
x\right. |\mathcal{B}_{k}|\left. x\right\rangle \left\langle x\right. |%
\mathcal{B}_{k}^{\ast }|\left. x\right\rangle .
\end{eqnarray*}

Applying the Hadamard (or element-wise or entry-wise) product (def.: $\left(
M\circ N\right) _{ab}=M_{ab}N_{ab}),$ e.g. $\mathcal{A}_{k}\circ \mathcal{A}%
_{k}^{\ast }=$ $\sum_{mn}\left( \mathcal{A}_{k}\circ \mathcal{A}_{k}^{\ast
}\right) _{mn}|\left. m\right\rangle \left\langle n\right.
|=\sum_{mn}\left\vert (\mathcal{A}_{k})_{mn}\right\vert ^{2}|\left.
m\right\rangle \left\langle n\right. |,$ we obtain

\begin{eqnarray*}
p^{(k)}(x,x|\theta ) &=&\left\langle x\right. |\mathcal{A}_{k}\circ \mathcal{%
A}_{k}^{\ast }|\left. x\right\rangle +\left\langle x\right. |\mathcal{B}%
_{k}\circ \mathcal{B}_{k}^{\ast }|\left. x\right\rangle  \\
&=&\left\langle x\right. |\left( \sum_{mn}\left( \left( \mathcal{A}_{k}\circ 
\mathcal{A}_{k}^{\ast }\right) _{mn}+\left( \mathcal{B}_{k}\circ \mathcal{B}%
_{k}^{\ast }\right) _{mn}|\left. m\right\rangle \left\langle n\right.
|\right) \right) |\left. x\right\rangle  \\
&=&\left\langle x\right. |\left( \sum_{mn}(\left\vert (\mathcal{A}%
_{k})_{mn}\right\vert ^{2}+\left\vert (\mathcal{B}_{k})_{mn}\right\vert
^{2})|\left. m\right\rangle \left\langle n\right. |\right) |\left.
x\right\rangle  \\
&=&\sum_{mn}(\left\vert (\mathcal{A}_{k})_{mn}\right\vert ^{2}+\left\vert (%
\mathcal{B}_{k})_{mn}\right\vert ^{2})\left\langle x\right. |\left.
m\right\rangle \left\langle n\right. |\left. x\right\rangle  \\
&=&\left\vert (\mathcal{A}_{k})_{xx}\right\vert ^{2}+\left\vert (\mathcal{B}%
_{k})_{xx}\right\vert ^{2},
\end{eqnarray*}%
or equivalently if $x=x=0,$ 
\begin{eqnarray*}
p^{(k)}(x,x|\theta ) &=&p^{(k)}\left( 0,0|\theta \right) \equiv
p_{0}^{(k)}(\theta ) \\
&=&\left\vert (\mathcal{A}_{k})_{00}\right\vert ^{2}+\left\vert (\mathcal{B}%
_{k})_{00}\right\vert ^{2}.
\end{eqnarray*}

Recall the $k$-th step Kraus generators $\mathcal{A}_{k}$ and $\mathcal{B}_{k}$ and their kernels (matrix elements in the continuous basis), ${A}_{k}$ and ${B}_{k}$ respectively, we obtain their expression in the discrete basis as
\[
\mathcal{A}_{k}=\sum_{mn}\left( \int_{0}^{2\pi }A_{k}(\phi ;\theta
)\left\langle m\right. |\left. \phi \right\rangle \left\langle \phi \right.
|\left. n\right\rangle d\phi \right) |\left. m\right\rangle \left\langle
n\right. | 
\]%
yields for the $00$-matrix elements 
\[
\left\vert (\mathcal{A}_{k})_{00}\right\vert ^{2}=\left\vert \int_{0}^{2\pi
}A_{k}(\phi ;\theta )d\phi \right\vert ^{2}. 
\]%
and similarly 
\[
\left\vert (\mathcal{B}_{k})_{00}\right\vert ^{2}=\left\vert \int_{0}^{2\pi
}B_{k}(\phi ;\theta )d\phi \right\vert ^{2}. 
\]

The likelihood function now reads 
\begin{eqnarray*}
\mathcal{L}_{n}^{(k)}(\theta |\left\{ d_{1}=0,...,d_{n}=0\right\} )
&=&\left( p_{0}^{(k)}(\theta )\right) ^{n} \\
&=&\left( \left\vert (\mathcal{A}_{k})_{00}\right\vert ^{2}+\left\vert (%
\mathcal{B}_{k})_{00}\right\vert ^{2}\right) ^{n} \\
&=&\left( \left\vert \int_{0}^{2\pi }A_{k}(\phi ;\theta )d\phi \right\vert
^{2}+\left\vert \int_{0}^{2\pi }B_{k}(\phi ;\theta )d\phi \right\vert
^{2}\right) ^{n}.
\end{eqnarray*}

The above integrals have been carried out analytically in Proposition 2, and the
likelihood function and its maximum has been studied and analyzed.

 \section{On the computational complexity of likelihood function}

    In this chapter we address the question of employing a QW for the task of estimation of a unknown parameter. In the estimation methodology of likelihood function the need of a large among of data is indispensable for the success of the estimation task. This requirement implies that a trading is set up between the size of gathered data and the number of operations needed for generating and collecting those data. In the concrete context of the QW-MLE framework the trading pair of actions corresponds to the number of CP maps on the walker systems density matrix generating the walker spread and a measure quantifying that spreading in terms of number of sites and their occupation probabilities. More specifically such a diffusion measure is the standard deviation of the position of a QWer for which the process of QW is know to exhibits a computation advantage in comparison to a classical random walk. In the biological processes of phylogenetic evolution, where to the problem of estimation is a central one and in which random processes are also employed, offers a paradigmatic case for possible use any quantum advantages, (see the related discussion in \cite{elljarv}). Next we provide an analytic discussion of the computation complexity of the likelihood function in the context of estimation and the advantages offered by QW-MLE developed so far.  
    
Recall that in a fixed number of steps $k$, a CRWer
diffuses over a range of order $\mathcal{O}(\sqrt{k}),$ while a QWer is
quadratically faster and diffuses in a range of order $\mathcal{O}(k).$ This
quadratic speed up implies that a MLE of a parameter via QW utilizes a more
extended set of points for the same number of steps, and in this way renders the QW based estimation
algorithm more effective in comparison with its classical counterpart. This well known
feature of the QW makes it an attractive process in applications (\cite{kempe} --\cite{kendon}).
Indeed we see that it also becomes an important feature  from the point of view of the estimation problem. Actually the
question of right size of the sample needed for the MLE to be successful for
the particular case of canonical densities and others is a problem of extensive discussion in various fields (see for example \cite{braunst}, \cite{taxa}).\\

The MLE based algorithm uses a set of $n$ QWs to build its likelihood
function. Therefore, a data box of size $k\times n$, representing the
number of steps $\times $ the numbers of QWs, constitutes the data resources
for the estimation problem in hand. Given that bigger data box is expected
to produce better estimations, the standard dilemma: 
\textit{larger number of steps
or larger number of QW?}, should be decided efficiently so that the
parameter $\theta $ is nearly optimal. Both in classical estimation theory
and in the present quantum likelihood estimation approach, $k$ \ and $n$ are
considered as quantities constituting two competing scarce resources. That is, the 
\textit{data box dilemma} transcribes in our context to then question: \textit{``fewer QWs running
for a longer time, or many QWs running for a shorter time?"}. Notice then the
significance of the quadratic speed up of QW (faster diffusion rate of a
QW), in connection with this \textit{data box dilemma}. A QRW produces $\mathcal{O}%
(k)\times n$ data points, in comparison to a CRW that could be used instead,
which produces $\mathcal{O}(\sqrt{k})\times n$ data points, so it is
expected to provide a more efficient (lower cost) parameter estimation, (see 
\cite{elljarv} for the \textit{data box dilemma} in the context of quantum
phylogenetics and \cite{taxa}, and references therein, for its original
form in classical phylogenetics).

\section{Discussion}
Quantum walks can be employed as devices for estimating unknown parameters.
The step operator of a QW on the integer lattice containing a local coin operator (chosen to be a single angle $\theta$ parametrized  $SO(2)$ matrix) and a non local conditional step operator, provide is a suitable framework for applying a maximum likelihood estimation technique for determining $\theta$. The QW dynamics is solved analytically for the case of arbitrary finite number of step. This is accomplished by determining the unitary evolution channel of the walk and its Kraus generators. The underlying algebraic structure manifested by the  Euclidean algebra is employed to show the translational invariance of the walk. That symmetry suggest the return probabilities of the walk describing closed loop events in the course of the evolution as a convenient probability mass function for building the likelihood function. The likelihood maximization of such reluctant walk is shown by combining analytic exacts results and some numerical investigation. Two important aspects of the those findings are presented: first, the operational aspect that shows how quantum projection measurements on the time evolved density matrix of the walker system can be used for computing the return probabilities and their associated likelihood function; second, complexity of evaluating the likelihood and the gathering of data. The data box dilemma is presented and  its importance for effective MLE is analyzed. It is argued that the known feature of a QW to manifest a quadratic speed up in the diffusion rate on the integer lattice in comparison with the corresponding rate of a classical random walk, offers a computational complexity advantage in employing a QW for the task of MLE of an unknown parameter. Some prospects for future extensions of this work would include the following items: the
extension of MLE via QW schemes to the case of multi-parameter coin matrices, further, 
the possibility of using QW schemes for estimating parameter of other gates or (CPTP) completely positive trace preserving maps of interest, since it is known that a QW can be regarded as a universal
computational primitive. Also QW-MLE for walks on other lattices and typologies would be a challenge (e.g. unpublished work by the authors that addresses the parameter estimation task in circular lattices for a unitary coin matrix provides interesting exact results).

\newpage













\section{Supplemental material: appendices}
\renewcommand{\thesubsection}{\Alph{subsection}}
\subsection{Chebyshev Polynomials - hypergeometric Functions $_{2}$F$%
_{1} $}

Recall some basic properties of Chebyshev orthogonal polynomials \cite%
{AS}. 
They are Chebyshev polynomials of first kind $T_{n}(x)~$and second
kind $U_{n}(x)$, of degree $n=0,1,2....$

The first kind polynomial $T_{n}\left( x\right) $ are defined recursively as
\ $T_{(n+1)}(x)=2xT_{n}(x)-T_{(n-1)}(x)$ and $T_{0}(x)=1,$ $T_{1}(x)=x,$ and
have a trigonometric definition $T_{n}(\cos \theta )=\cos (n\theta ).$

The second kind of polynomial $U_{n}\left( x\right) $ are defined
recursively as $U_{(n+1)}(x)=2xU_{n}(x)-U_{(n-1)}(x)$ and $U_{0}(x)=1,$ \ $%
U_{1}(x)=2x$ and have a trigonometric definition $\ U_{n}(\cos \theta )=%
\frac{\sin ((n+1)\theta )}{\sin \theta }.$\\

\noindent
\textbf{Lemma} : Chebyshev polynomials of second kind $U_{k}(\cos \phi )$
wrt $\cos \phi ,$ and those with scaled argument i.e. $U_{k}(\xi
):=U_{k}(\lambda (\theta )\cos \phi ),$\ where $\lambda (\theta )=\cos
\theta $ is a function of parameter $\theta ,$ satisfy the following
relations regarding the derivation and integration 
\begin{eqnarray*}
\frac{\partial }{\partial \theta }\ \left[ U_{2r+1}(\xi ){}\right] &=&-\tan
\theta \sum_{m:odd}^{2r+1}m[\cos \phi U_{m-2}(\xi )+U_{m}(\xi ){}], \\
\frac{\partial }{\partial \theta }\ \left[ U_{2r}(\xi ){}\right] &=&-\tan
\theta \sum_{m:even}^{2r}m[U_{m-2}(\xi ){}+U_{m}(\xi ){}],
\end{eqnarray*}%
\begin{equation*}
\int\limits_{0}^{2\pi }{}\left\{ U_{2r+1}(\lambda (\theta )\cos \phi
)\right\} {}\frac{d\phi }{2\pi }=0,\text{ \ \ \ \ \ \ \ \ }%
\int\limits_{0}^{2\pi }{}\left\{ U_{2r}(\lambda (\theta )\cos \phi
){}\right\} \frac{d\phi }{2\pi }=Y_{2r}(\lambda (\theta )),
\end{equation*}%
as well as
\begin{eqnarray*}
2\int\limits_{0}^{2\pi }\left\{ \cos \phi U_{2r}(\lambda (\theta )\cos \phi
)\right\} {}\frac{d\phi }{2\pi } &=&\int\limits_{0}^{2\pi }\left\{
U_{2r-1}(\lambda (\theta )\cos \phi )\right\} {}\frac{d\phi }{2\pi } \\
+\int\limits_{0}^{2\pi }\left\{ U_{2r+1}(\lambda (\theta )\cos \phi
)\right\} {}\frac{d\phi }{2\pi } &=&0,
\end{eqnarray*}%
and similarly 
\begin{eqnarray*}
&&2\int\limits_{0}^{2\pi }\left\{ \cos \phi U_{2r+1}(\lambda (\theta )\cos
\phi )\right\} {}\frac{d\phi }{2\pi }= 
\int_{0}^{2\pi }\left\{ U_{2r}(\lambda (\theta )\cos \phi
)\right\} {}\frac{d\phi }{2\pi }+\int\limits_{0}^{2\pi }\left\{
U_{2r+2}(\lambda (\theta )\cos \phi )\right\} {}\frac{d\phi }{2\pi } \\[.2cm]
&=:&Y_{2r}(\lambda (\theta ))+Y_{2r+2}(\lambda (\theta )).
\end{eqnarray*}
\mbox{}\hfill$\Box$\\
\noindent
The proof is based on the definition of polynomials
\begin{equation*}
U_{n}(x)=\sum_{l=0}^{%
\lfloor n/2\rfloor }(-1)^{l}\binom{n-l}{l}(2x)^{n-2l},n>0, \end{equation*} and their
defining recurrence relations. The explicitly evaluation of integral 
\begin{equation*}
Y_{m}(\lambda (\theta ))=\int\limits_{0}^{2\pi }{}U_{m}(\lambda (\theta
)\cos \phi ){}\frac{d\phi }{2\pi },
\end{equation*}
provides the definition of $Y_{m}$
polynomials: $Y_{m=2k+1}(\lambda )=0,$ and 
\begin{equation*}
Y_{m=2k}(\lambda (\theta ))=\frac{1}{\pi }\sum_{l=0}^{\lfloor m/2\rfloor
}c_{l}(\theta )\frac{(m-2l-1)!!}{\left( m-2l\right) !!},
\end{equation*}%
where \ $c_{l}(\theta )=(-1)^{l}\binom{m-l}{l}(2\cos \theta )^{m-2l}.$

\subsection{Euclidean algebra}

\noindent
\textbf{Euclidean algebra}\\
Let the operator position $L$ $\ $and the step
operators $E_{\pm }$ satisfying the commutation relations of the Euclidean
algebra $E(2)\equiv ISO\left( 2\right) =span\left\{ L,E_{+},E_{-}\right\} $
viz. $[L,E_{\pm }]=\pm E_{\pm }$ \ and $[E_{+},E_{-}]=0.$ If the phase
operator $\widehat{\Phi }$ is defined from the relation $E_{\pm }=e^{\pm i%
\widehat{\Phi }},$ then the phase-position operators satisfy the canonical
algebra $[L,\widehat{\Phi }]=i\mathbb{I}\mathbf{.}$ \ 

There are two irreducible representations of Euclidean algebra generators
carried out by vectors spaces $\mathcal{H}_{w}$ and $\mathcal{H}_{w}^{\ast }$%
which are respectively the eigenspace of position operator $L$ and of
unitary step operators $E_{\pm }$ as well as of Hermitian phase operator $%
\widehat{\Phi }.$ The latter are inter-related as $E_{\pm }=e^{\pm i\widehat{%
\Phi }}.$

Explicitly the two vector spaces are: the walker Hilbert space $\mathcal{H}%
_{w}=l_{2}\left( 
\mathbb{Z}
\right) $ and its discrete spanning set$\ $of vectors $span\left \{
\left
\vert m\right \rangle \right \} _{m\in 
\mathbb{Z}
},$ forming an orthogonal viz. $\left \langle m\right. \left \vert m^{\prime
}\right \rangle =\delta _{mm^{\prime }}$ and complete viz. $\left.\sum
\right._{m\in 
\mathbb{Z}
}\left \vert m\right \rangle \left \langle m\right \vert =\mathbb{I}_{H_{w}}$
basis, and its dual space $\mathcal{H}_{w}^{\ast }=L_{2}\left( [0,2\pi
\right) ,\frac{d\phi }{2\pi })$ with its continuous spanning set of vectors $%
\{ \left \vert \phi \right \rangle ;\phi \in \lbrack 0,2\pi ),\frac{d\phi }{%
2\pi }\},$ forming a generalized orthogonal viz. \ $\left \langle \phi |\phi
^{\prime }\right \rangle =\delta (\phi -\phi ^{\prime })$ and complete viz. $%
\left.\int\right._{0}^{2\pi }\left \vert \phi \right \rangle \left \langle \phi
\right \vert \frac{d\phi }{2\pi }=\mathbb{I}_{H_{w}^{\ast }},$ basis,
inter-related as $\left \vert \phi \right \rangle =\left.\sum
\right._{m\in 
\mathbb{Z}
}e^{i\phi m}\left \vert m\right \rangle $ and $\left \vert m\right \rangle
=\int_{0}^{2\pi }e^{-i\phi m}\left \vert \phi \right \rangle \frac{d\phi }{%
2\pi }.$

The respective differential representations of the generators in the
continuous basis are$\ \ E_{\pm }\left\vert \phi \right\rangle =e^{\pm i\phi
}\left\vert \phi \right\rangle $ and\ $L\left\vert \phi \right\rangle =\frac{%
1}{i}\frac{\partial }{\partial \phi }\left\vert \phi \right\rangle ,$ and
the corresponding one in the discrete basis are $E_{\pm }\left\vert
m\right\rangle =\left\vert m\pm 1\right\rangle $ and\ $L\left\vert
m\right\rangle =m\left\vert m\right\rangle .$\\

\end{document}